\documentclass[fleqn,usenatbib]{mnras}

\usepackage{newtxtext,newtxmath}

\usepackage[T1]{fontenc}

\DeclareRobustCommand{\VAN}[3]{#2}
\let\VANthebibliography\thebibliography
\def\thebibliography{\DeclareRobustCommand{\VAN}[3]{##3}\VANthebibliography}

\usepackage{graphicx}	
\usepackage{amsmath}	
\usepackage{amssymb}	
\usepackage{xcolor}

\def\rxj{RX\,J0209.6-7427}
\def\flux{~ergs~cm$^{-2}$~s$^{-1}$}
\def\luminosity{~ergs~s$^{-1}$}

\title[\rxj]{A major optical \& X-ray outburst from the Magellanic Bridge source \rxj}

\author[M. J. Coe et al.]{
M.~J. Coe,$^{1}$\thanks{E-mail: mjcoe@soton.ac.uk}
I.~M. Monageng,$^{2,3}$
E.~S. Bartlett,$^{4,5}$
D.~A.~H. Buckley,$^{2}$
and A. Udalski$^{6}$
\\
$^{1}$Physics \& Astronomy, The University of Southampton, SO17 1BJ, UK\\
$^{2}$South African Astronomical Observatory, P.O Box 9, Observatory, 7935, Cape Town, South Africa\\
$^{3}$Department of Astronomy, University of Cape Town, Private Bag X3, Rondebosch 7701, South Africa\\
$^{4}$ESO - European Southern Observatory, Alonso de C{\' o}rdova 3107, Vitacura, Casilla 19001, Santiago de Chile, Chile\\
$^{5}$UK Astronomy Technology Centre, Royal Observatory, Blackford Hill, Edinburgh, EH9 3HJ, UK\\
$^{6}$ Astronomical Observatory, University of Warsaw, Al. Ujazdowskie 4, 00-478 Warszawa, Poland \\
}

\date{Accepted 2020 March 20. Received 2020 March 20; in original form 2020 February 24}

\pubyear{2020}

\begin{document}
\label{firstpage}
\pagerange{\pageref{firstpage}--\pageref{lastpage}}
\maketitle

\begin{abstract}
\rxj{} is an X-ray source in the Magellanic Bridge that was first detected in 1993, but not seen again till 2019. It has been identified as a member of the Be/X-ray binary class, a category of objects that are well established as bright, often-unpredictable transients. Such systems are rarely known in the Bridge, possibly because they lie outside the area most commonly studied by X-ray telescopes. Whatever the reason for the sparse number of such systems in the Bridge, they can provide useful tools for trying to understand the result of the tidal dynamics of the two Magellanic Clouds. In this paper the nature of the object is explored with the help of new data obtained during the latest outburst. In particular, the first optical spectrum of the counterpart is presented to help classify the star, plus measurements of the Balmer emission lines over several years are used to investigate changes in the size and structure of the circumstellar disk.
\end{abstract}




\section{Introduction}

The Magellanic Bridge is the name given to the region linking the two Magellanic Clouds. It is clearly visible in HI radio mapping \citep{Putman2000} and in young stellar population \citep{skowron2014,jd2020}, and is thought to represent material torn out of the Small Magellanic Cloud (SMC) as a result of severe tidal interactions between the two galaxies. The age and content of this material is not well understood and identifying the stellar population is a valuable tool. In particular, the study of High Mass X-ray Binaries (HMXBs) provides a valuable insight in star formation and evolution because of their relatively short life times \citep{Grimm2003}. To date there are only 4 confirmed HMXBs in the Bridge in contrast to the $\sim$80 in the nearby SMC \citep{Coe2015}. This small number may, in part, be due to the limited X-ray mapping of this region, but it may also be providing some important insight into the physical nature of the Bridge. In this paper we report on a relatively rare event - a super-Eddington outburst from one of the 4 known HMXB systems.

\rxj{} was originally reported by \cite{Kahabka2005} using two observations from 1993 with the {\it ROSAT} observatory. \textcolor{black}{The peak X-ray luminosity in the 0.1 - 2.4 keV range was reported in 1993 as $1.0 \times 10^{38}$\luminosity{}} assuming an SMC distance of 60 kpc \citep{Kahabka2005}. Those authors re-analysed the original data and located the source to 12" and then carried out an optical study of the region, suggesting that a previously-unclassified OB star with a type B0 - B1V was the counterpart. They measured the red spectrum of this star and determined it exhibited H$\alpha$ emission with an Equivalent Width value of -10.8$\pm$0.2\AA.

In addition \cite{Kahabka2005} carried out a timing analysis of the X-ray flux and suggested that the strong period they saw at $\sim$39d might be the binary period of the system. But they also warned that the separation of the two observations ($\sim$200d) might be influencing the result.

Subsequently the source remained undetected at X-ray wavelengths until it was picked up by the Monitor of All-sky X-ray Image telescope (MAXI) on 20 November 2019 at the start of a new outburst \citep{Negoro2019}. They initially named the source MAXI J0206-749 as the large uncertainty in the position determined from MAXI did not immediately link it to RX J0209.6-7427. That link was made by observations from the {\it Swift} observatory \citep{Kennea2019} which scanned the area around the MAXI position and only detected one new outbursting object - \rxj{}.

Follow-up X-ray observations by the Neutron Star Interior Composition Explorer (NICER) quickly revealed the detection of a previously unknown pulse period of 9.29s \citep{Iwakiri2019}.\textcolor{black}{ The combination of the MAXI and NICER data suggests that the peak luminosity from this outburst in the photon energy range 0.2 - 10 keV reached in excess of $ 10^{39}$\luminosity{}.}

In this paper we report on optical spectroscopy and photometry of the counterpart to \rxj{} contemporaneous with the 2019 X-ray outburst, including the first blue spectrum of the optical counterpart. In addition we report on historical optical data (photometric and spectroscopic) showing the behaviour of the counterpart over the last $\sim$15 years.

\section{Observations}

\subsection{MAXI}

\begin{figure}

	\includegraphics[width=9cm,angle=-0]{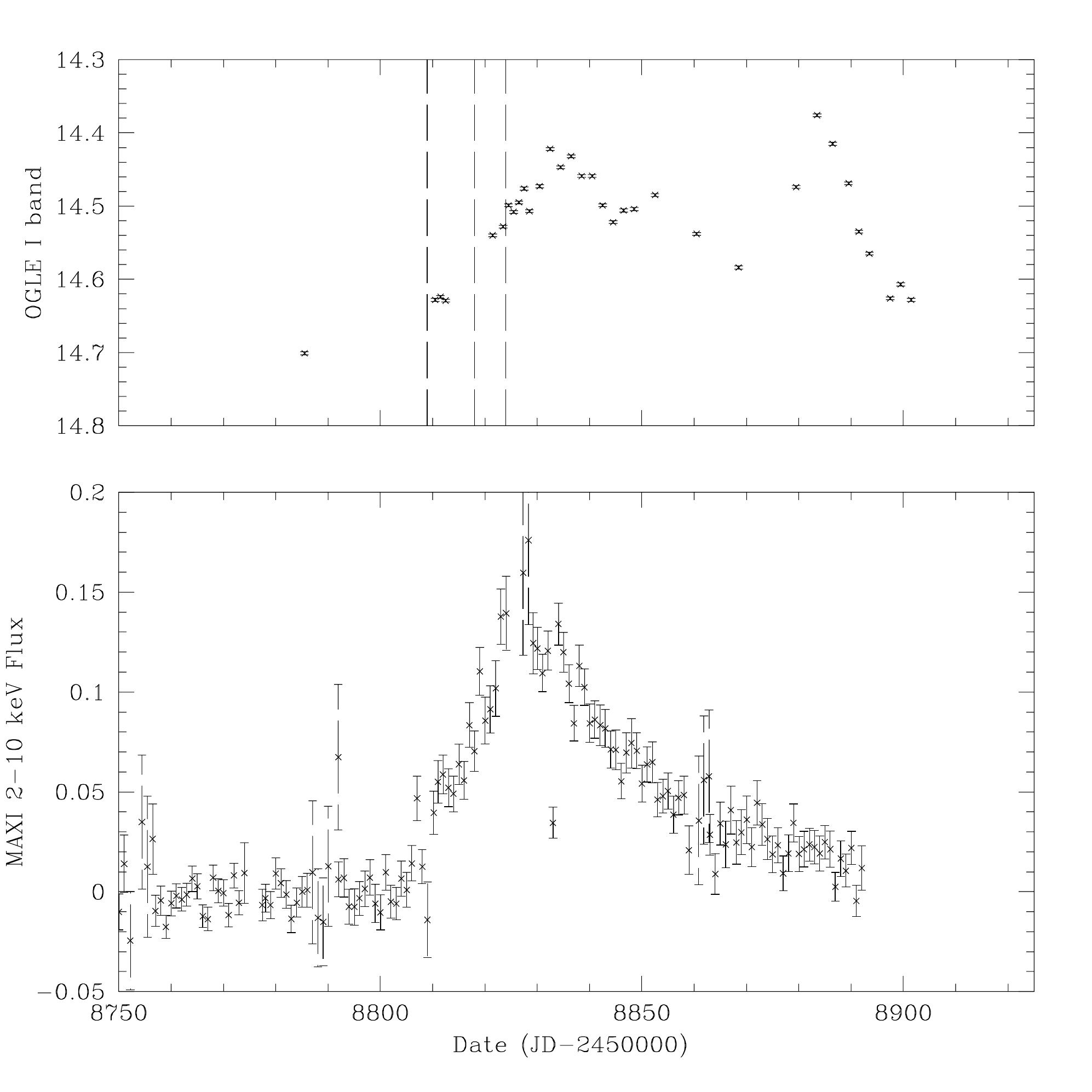}
    \caption{Comparison between the OGLE I-band brightness (\textcolor{black}{ in magnitudes} - upper panel) and the 2-10 keV data from MAXI (\textcolor{black}{ in instrumental counts/s} - lower panel). The dates of the SALT observations are shown in the upper panel by the vertical dashed lines.}
    \label{fig:max_ogle}
\end{figure}

The new outburst from this source was first reported by \cite{Negoro2019} using observations made from the MAXI instrument on the ISS \citep{Matsuoka2009}. The instrument continued to monitor the source throughout the outburst and the resulting public 2-10 keV X-ray lightcurve is shown in the lower panel of Fig \ref{fig:max_ogle}.

\subsection{OGLE}

The OGLE project \citep{Udalski1997}  provides long term I-band photometry with a cadence of 1-3 days. Once the optical counterpart of the source had been identified through the accurate Swift XRT position \citep{Kennea2019} with a previously known Be star \citep{Kahabka2005}, it was possible to retrieve many years of photometric monitoring from OGLE IV in the I-band, plus sparser coverage in the V band. The source is identified as MBR109.10.389D  in the I band and MBR109.10.V.1488 in the V band. 

\begin{figure}

	\includegraphics[width=9cm,angle=-00]{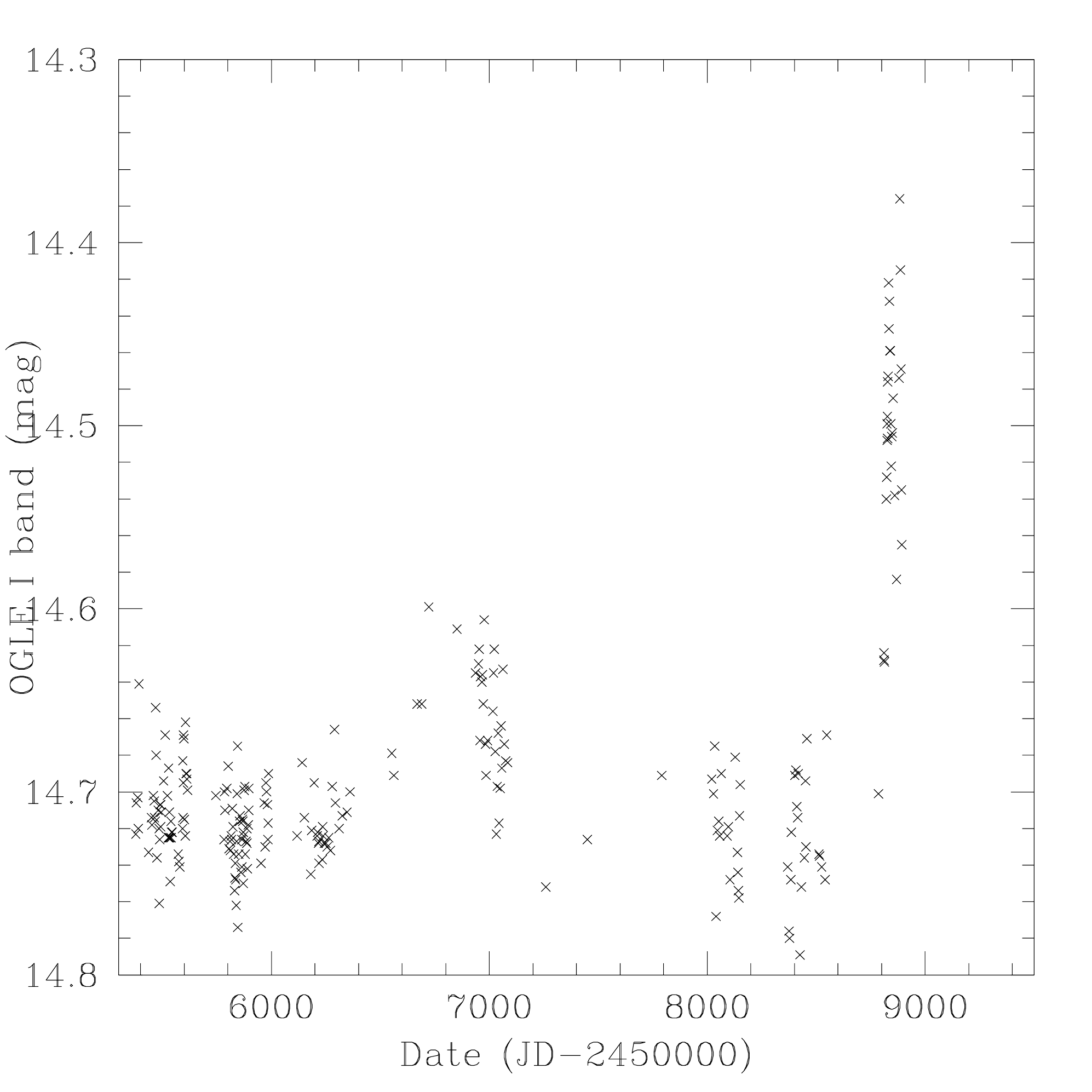}
    \caption{All the OGLE I-band data from the last$\sim$10 years.}
    \label{fig:ogle}
\end{figure}

The subset of the OGLE I-band data that coincide with the current X-ray outburst are shown in the top panel of Fig \ref{fig:max_ogle}. All the OGLE I-band data are presented in Fig \ref{fig:ogle}.

The OGLE data were searched for any evidence of a periodicity that might be associated with a binary period. The Corbet diagram for sources of this nature would suggest that the binary period lies in the range 20 - 30d. So, the data with, and without, the current outburst were searched using Lomb-Scargle routines, but revealed no significant periods in the search range of 2 - 100d.   

\subsection{SAAO 1.9m telescope}

\begin{figure}
\centering
\includegraphics[width=0.45\textwidth]{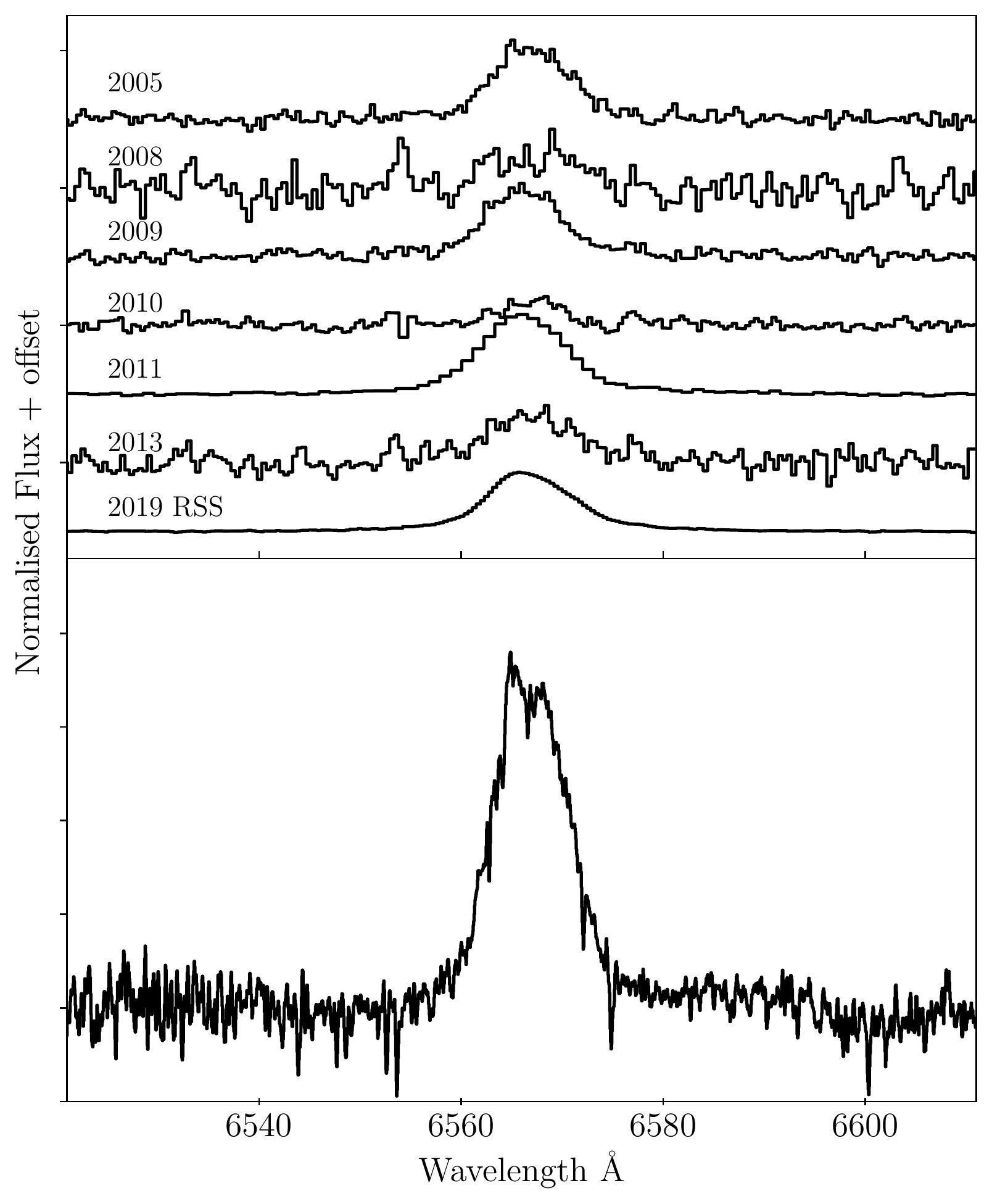}
\caption{\emph{Top panel:} Normalised H$\alpha$ spectra of \rxj{} from SAAO 1.9m and EFOSC; also including the RSS spectrum taken by SALT in 2019. \emph{Bottom panel:} The highest resolution spectrum taken 6 Dec 2019 with the HRS instrument on SALT. The lower spectrum in this panel is the same data slightly smoothed with a Boxcar function with a width 0.213\AA.}
\label{fig:ha}
\end{figure}

Optical spectra of optical counterpart to \rxj{} were obtained over several years with the South Africa Astronomical Observatory (SAAO) 1.9-m telescope, Sutherland Observatory, South Africa. The spectrograph  was used  with the SITe detector and a 1200 l/mm grating over the range 6300-6900\AA. The dispersion was 1.0\AA/pixel and the resulting SNR approximately 10. The spectra are shown in Fig \ref{fig:ha} and the determined Equivalent Width measurements for the H$\alpha$ emission line are presented in Table \ref{tab:ha}.

\subsection{SALT}
We observed the optical counterpart of \rxj{} with the Southern African Large Telescope (SALT; \citealt{Buckley2006}) using the Robert Stobie Spectrograph (RSS; \citealt{2003SPIE.4841.1463B,2003SPIE.4841.1634K}) and the High Resolution Spectrograph (HRS; \citealt{2010SPIE.7735E..4FB,2014SPIE.9147E..6TC}). The observation performed with the RSS was obtained on 21 November 2019 with grating PG1800 (resolution $\sim$2.6 \AA) and covers a wavelength range $6000-7250$~\AA. An exposure time of 1600~s was used for the observation. The primary steps of data reduction were performed with the SALT pipeline \citep{2012ascl.soft07010C}, which include overscan correction, bias subtraction, gain correction and amplifier cross-talk correction. The remaining steps were carried out in a standard way using \textsc{iraf}\footnote{Image Reduction and Analysis Facility: iraf.noao.edu} (arc line identification, background subtraction and extraction of the 1D spectrum).\\
The two observations obtained with the HRS were performed on 30 November 2019 and 6 December 2019 using the low resolution ($R \sim 14000$) and medium resolution ($R \sim 40000$) modes, respectively. We used exposure times of 1800~s and 2200~s for the low resolution and medium resolution mode observations, respectively. The HRS observations cover a wavelength range of $5500-8800$~\AA. The SALT pipeline \citep{2015ascl.soft11005C} was used for the primary reductions, while the remainder of the reductions were done using the \textsc{midas feros} \citep{1999ASPC..188..331S} and \textsc{echelle} \citep{1992ESOC...41..177B} packages (which include background subtraction, identification of arc lines, blaze function removal and order merging). See \cite{2016MNRAS.459.3068K} for a full description of the reduction procedure.  The resulting spectra are shown in Fig \ref{fig:ha}.

\subsection{NTT/EFOSC2 data}

\rxj\ has also been observed with the ESO Faint Object Spectrograph and Camera v2. \citep[EFOSC2][]{Buzzoni1984}, mounted at the Nasmyth B focus of the 3.6m New Technology Telescope (NTT) at La Silla Observatory, Chile, on the night of 2011 December 11. At that time the OGLE data showed the source to be at I=14.7 compared to the SALT observations mentioned above when the source was somewhat brighter at I=14.4.The instrument was in longslit mode with a slit width of 1.5 arcsec and instrument binning 2$\times$2. Grisms 14 and 20 were used to obtain spectra at both blue ($\sim$3500 - 5000 \AA{}) and red ($\sim$6000 - 7000\AA{}) wavelengths i.e. covering all the Balmer transitions. For Grism 14, this lead to a dispersion of $\sim2$\AA{}/pix and a resolution of $\sim11$\AA{}/fwhm. For Grism 20, this lead to a dispersion of $\sim$1\AA{}/pix and a resolution of $\sim$6\AA{}/fwhm. Both spectra have a SNR of $\sim70$. The spectra were reduced, extracted and calibrated using the standard \textsc{iraf} packages.

\begin{figure*}
\begin{center}
    \includegraphics{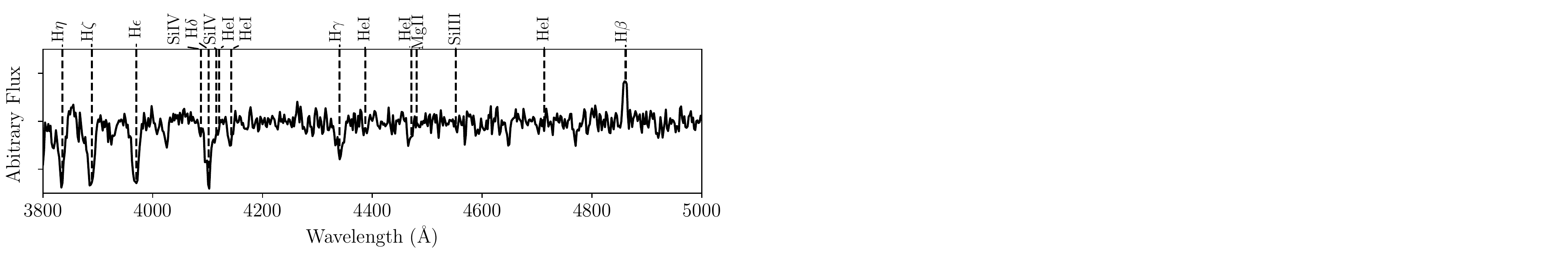}
    \caption{3800 - 5000 \AA{} spectrum from EFOSC2. A radial velocity correction of $\sim$-150 km~s$^{-1}$ has been applied. Line transitions discussed in the text are marked.}
    \label{fig:blue}
    \end{center}
\end{figure*}

Shown in Fig ~\ref{fig:blue} is the blue spectrum obtained from EFOSC2. The spectrum has had a radial velocity correction applied corresponding to $\sim$-150 km~s$^{-1}$, derived from a Voigt profile fit to the H$\alpha$ profile, taken immediately after. The H$\beta$ Balmer line (4861\AA) is in emission like H$\alpha$, but the rest of the Balmer series show absorption profiles.

\begin{table*}
	\centering
	\caption{Summary of \textbf{observations of \rxj{} in the H$\alpha$ wavelength range. The errors on the equivalent width measurements are calculated using equation (7) of \citet{Vollmann2006}}}
	\label{tab:ha}
	\begin{tabular}{lccccl} 

Date&Julian Date&Telescope&SNR&EW&Comments \\
\hline
26 July 2004 & 2453213.45 & ESO/VLT&$\sim 200$ &-- 10.8 $\pm$ 0.2 &\cite{Kahabka2005} \\
30 Oct 2005& 2453673.62&SAAO 1.9m& 13.5 &--10.$\pm$1.4& Good strong spectrum, possibly a flat top\\
5 Dec 2008& 2454805.58&SAAO1.9m& 5.0 &$\le$--10.9& Weak SNR, shape unclear if there at all \\
11 Dec 2009& 2455177.46&SAAO1.9m& 16.5 &--9.7$\pm$1.2& Strong signal, single peak \\
21 Dec 2010& 2455551.58&SAAO1.9m& 20.1 &--3.0$\pm$1.5& Weak \\
11 Dec 2011&2455907.45&ESO/NTT& 70.5 &--12.5$\pm1.2$&strong single peak \\
3 Nov 2013& 2456599.62&SAAO1.9m& 8.5 &--8.1$\pm$2.6&Good SNR, single peak \\
21 Nov 2019&2458809.35&SALT& 197.5 & --9.3$\pm$0.1& Strong single peak\\
30 Nov 2019&2458818.33&SALT& 16.2 &--8.9$\pm$0.1& Strong single peak \\
6 Dec 2019&2458824.29&SALT& 16.8 &--8.2$\pm$1.3&  Evidence for a flat top or double peak \\


		\hline
	\end{tabular}
\end{table*}

\section{Discussion}

\subsection{Circumstellar disk size}


If the Keplerian velocity distribution of matter in the disk can be assumed then \cite{Huang1972} demonstrated that the size of the H$\alpha$ region of the circumstellar disk can be estimated by using the peak separation, $\Delta$V, of the double peaked H$\alpha$ emission lines.

From the spectrum shown in the lower panel of Fig \ref{fig:ha} 
the measured peak separation is 2.4$\pm$0.1 \AA,  or, in velocity space, 110$\pm$5 km/s. \textcolor{black}{The inclination angle of the disk to the line of sight, {\it i}, is unknown and clearly affects the interpretation of any rotational velocities. So it is assumed that {\it sin i = 0.5} for what follows. We also assume that the stellar classification lies in the range B0IV - B2V i.e. with a stellar radius size range of 12.1$R_\odot$ - 5.6$R_\odot$, and stellar masses of 23 - 11$M_\odot$ respectively. Inputting those numbers for {\it sin i} and the stellar radii into Eq 1 of \cite{Monageng2017}, the measured peak separation suggests an H$\alpha$ emitting disk of radius 130-280 $R_\odot$ or $(0.9-1.9) \times 10^{11}$m. }

There is a well-established relationship between the H$\alpha$ EW and the predicted peak separation - see \cite{Zamanov2001} for applications to a sample of typical Be stars found in X-ray binaries. Taking the H$\alpha$ typical value during the recent outburst as -9$\pm1$\AA ~ this predicts a peak separation of 96$\pm3$ km/s - very similar to the value of 110$\pm$5  km/s directly measured above. 

This value for the H$\alpha$ disc size is at the top end of the range seen for other similar systems in X-ray active states - see \cite{Monageng2017} for a sample of such objects. But we note that this outburst is an extreme one and hence the disk size is also probably unusually large.

\subsection{Circumstellar disk variability}

\textcolor{black}{From Fig \ref{fig:max_ogle} it can be seen that the X-ray brightening correlates remarkably well with that of the optical I-band data. This is clearly a rare event as we can see from Fig \ref{fig:ogle} that the optical emission has not been this bright for more than a decade. Presumably this brightening is triggered by an exceptional mass ejection from the B-type star. Had the disk grown gradually then the X-ray emission would have started to appear at a low level around periastron passage of the neutron star - the so-called Type I outbursts. But in this case the inferred extreme mass outflow has clearly filled the entire orbital space triggering a large and extensive X-ray outburst. This is typical of a Type II outburst.}

Though the OGLE project primarily collects data in the I-band, it occasionally also takes V-band measurements. This makes it possible to construct a colour-magnitude diagram (CMD) over time. In the past such diagrams have been very helpful in displaying the disk colour changes during growth and decline by revealing a looping pattern on the CMD - see \cite{Rajoelimanana2011} for several examples of such behaviour. The extent of the looping is believed to be related to the inclination of the disk to our line of sight. 

If no loop is evident, then there is normally a strong correlation between the disk colour and the I band magnitude. In general, disks reveal increasing red colours the larger they get. This is because the outer sections of the disk are always cooler than the stellar temperature - see \cite{Monageng2019} for an excellent example of such behaviour in SXP 91.1.

\begin{figure}
	\includegraphics[width=8cm,angle=-00]{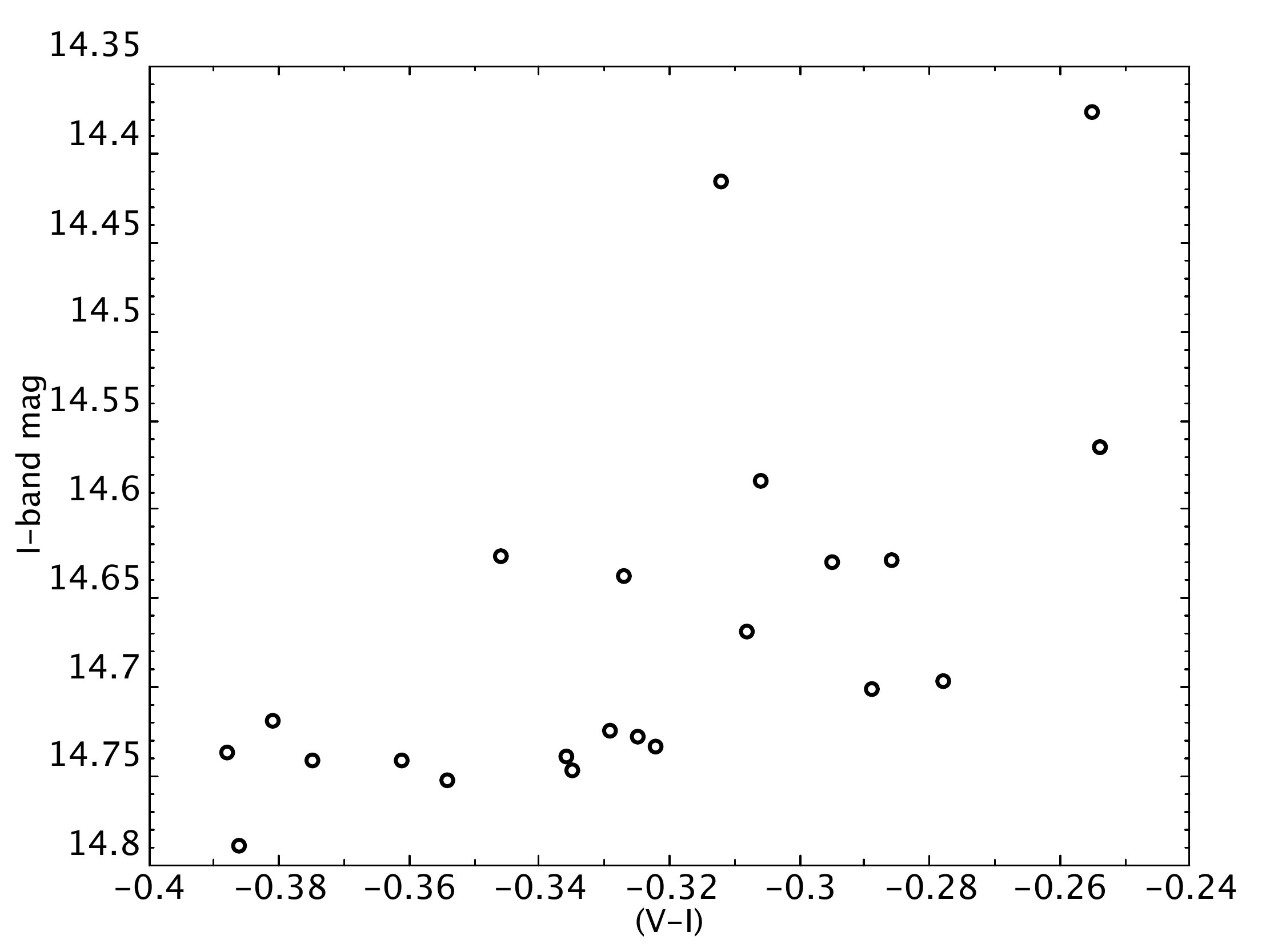}
    \caption{Colour magnitude plot from OGLE data.}
    \label{fig:cmd}
\end{figure}

However in this case the behaviour pattern is more complex - see Figure ~\ref{fig:cmd}. What can be said is that the most recent, brightest points show up in the upper right quadrant of this colour-magnitude change. Prior to that there have been significant smaller colour changes, but with little change in brightness.  It could be a system that normally goes through erratic variability, where the disc does not completely dissipate before renewed disc-growth episodes occur. Certainly the detection of the H$\alpha$ line always in emission seems to support this idea.  

The range in colour in Fig.~\ref{fig:cmd} is on the order of $\sim$0.1~mag, which is typically seen during disc-growth in other BeXBs (see \citealt{Rajoelimanana2011}). In the case of \rxj{}, however, the change in brightness from the Be disc is relatively small. From the H-alpha profiles in Figs.~\ref{fig:ha} 
we can infer a low viewing angle of the disc (i.e. close to face-on) \citep{hm1986}, which would in turn result in a strong correlation between the colour and magnitude \citep{1983HvaOB...7...55H,Rajoelimanana2011}. The historic relatively small change in magnitude during disc growth is possibly an indicator of an optically thin disc which does not contribute too much to the continuum emission for the brightness to change significantly.  

Finally, we note that the current outburst reveals two strong optical peaks separated by $\sim$50d - see Fig \ref{fig:max_ogle}. This could be indicative of a binary period, with the already enlarged circumstellar disk being further extended at the time of periastron by interaction with the companion neutron star. Alternatively it could simply be evidence for a system showing multiple mass ejections during this very active period.

\subsection{Spectral type of the optical counterpart}

Classifying B-type stars in low metallicity environments, such as the SMC, is not straight forward; the metal lines that are primarily used as a diagnostic are noticeably weaker \citep[see][for a detailed discussion]{Lennon1997}. This is compounded when looking at low luminosity stars, as the strengths of these lines also correlate with luminosity \citep[see e.g.][]{Evans2004}. As such it is difficult to determine whether a metal line is not present due to temperature, or just not visible above the noise. High resolution, high signal to noise spectra are required to provide an accurate spectral classification. Nevertheless, \citet{Evans2004} provide a classification criteria for B-type stars in the SMC which we adopt here in a rough attempt at a classification. There appears to be evidence for Si\,{\sc ii}\,$\lambda 4088$ which would point to a classification B1 or later. There is no evidence for Mg\,{\sc ii}$\lambda 4481$ and a possible feature at $\lambda 4553$ which, if real, would correspond to Si\,{\sc iii}. This would constrain the spectral type to B2 or earlier. It is not possible to say whether Si\,{\sc ii}\,$\lambda 4116$ is absent or just not visible above the noise, which would further constrain the spectral type. 

Therefore we conclude that the spectral range proposed by \cite{Kahabka2005} - based upon their photometry - of B0.5 to B1.5 is consistent with our own conclusions of B1 - B2. Higher quality spectroscopic data are required to narrow down the range more precisely.Nonetheless a counterpart with a spectral class in the B1-B2 range would be completely consistent with the spectral types seen in many other Be/X-ray binary systems in the SMC region \citep{McBride2008}.

\subsection{Comparing 1993 outbursts to that of 2019}

\begin{figure}

	\includegraphics[width=9cm,angle=-00]{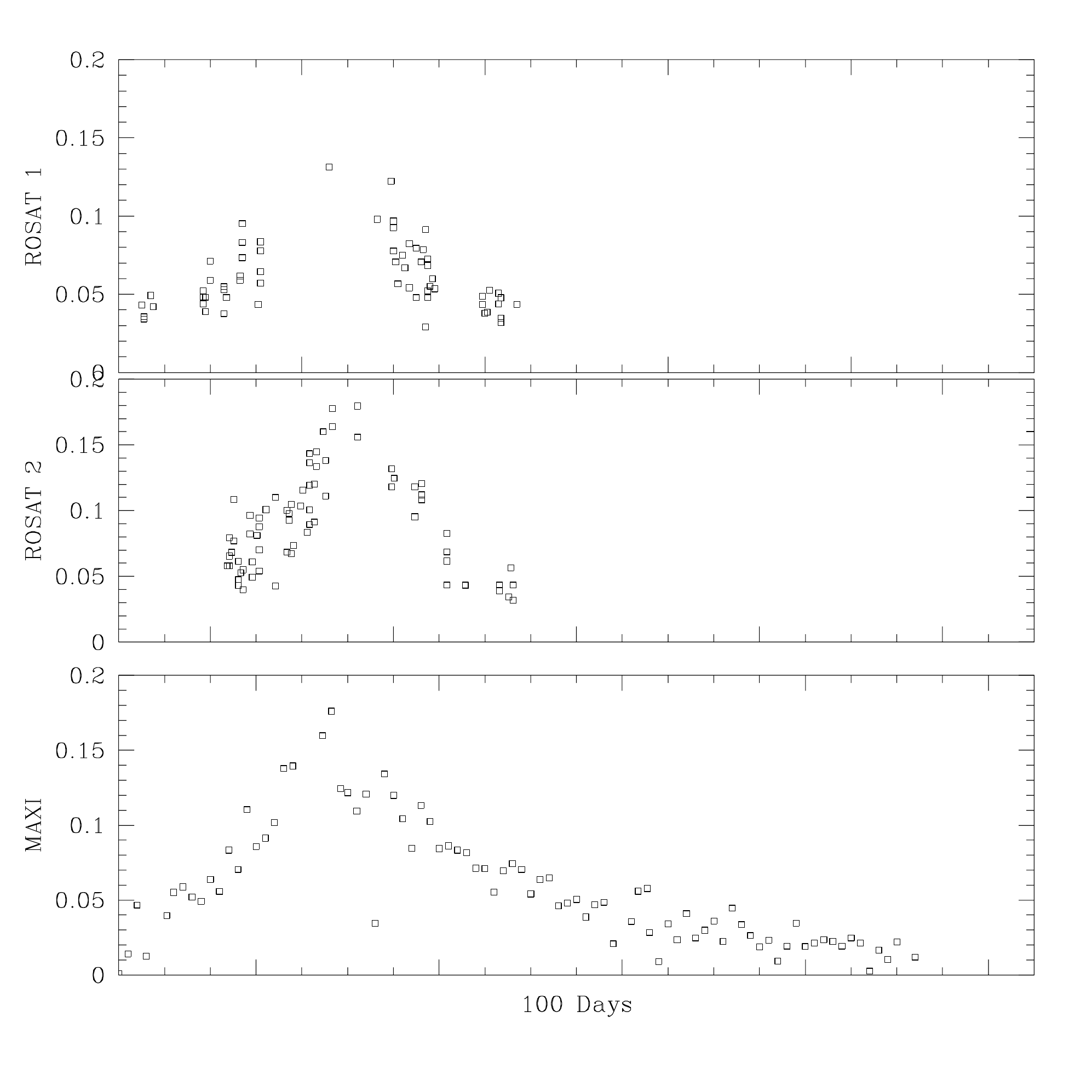}
    \caption{Comparison of the X-ray profiles of the two outbursts from 1993 \citep{Kahabka2005} (top 2 panels) with the 2019 one. The scale of the time axes are the same in all three cases and extend over 100d.}
    \label{fig:rm}
\end{figure}

In Figure ~\ref{fig:rm} the historical X-ray outburst profiles from 1993 \citep{Kahabka2005} are compared to the X-ray profile from the outburst being discussed in this work. All three outbursts are shown in this figure on the same timescale. The two 1993 outbursts were detected for $\sim$50d whereas the 2019 outburst is closer to $\sim$100d. The main differences seem to be longer rise time in 2019 plus a much longer decay time after the peak. The exponential rise time of the second 1993 outburst was $\sim$10d, whereas it is more like $\sim$15d in 2019. The longer decay time in 2019 may be, at least, partially attributable to the second optical outburst seen in the OGLE data which could have rejuvenated the circumstellar disk as it was beginning to decline. Or it could be due to some binary interaction. All of this provides valuable constraints for any future modelling of the mass injection rates into the disk and its consequential growth and decay. Previous Smooth Particle Hydrodynamic modelling of super-Eddington X-ray outbursts in a Be/X-ray system has shown that this can be quite a complex process with many changes in mass injection rates during the outburst \citep{brown2019}.

The peak X-ray luminosity reported in 1993 was $1.0 \times 10^{38}$\luminosity{} assuming an SMC distance of 60 kpc \citep{Kahabka2005}. NICER reported a flux of $3.7 \times 10^{-10}$ \flux{} on 21 November 2019 \citep[TJD 58808][]{Iwakiri2019}. This corresponds to a luminosity of $1.7 \times 10^{38}$\luminosity{}. At that time the MAXI flux was 0.008 ct/s, which subsequently increased to a peak of 0.14 ct/s. Scaling the NICER flux by this increase (a factor of 17.5) suggests that the peak luminosity from this outburst reached in excess of $ 10^{39}$\luminosity{} for a source in the SMC - clearly super-Eddington in nature.

\textcolor{black}{Outbursts of such a magnitude are certainly not common in the Small Magellanic Cloud, with only one previous example being documented in the last decade or so - that of SMC X-3 \citep{Townsend2017}. That source peaked at $1.2 \times 10^{39}$\luminosity{}, very similar to what is reported here for \rxj{}. It also exhibited a rare high I-band brightness. Clearly such exceptional X-ray outbursts are driven by extreme behaviour in the mass donor, Be star, and the X-ray channel provides an excellent monitoring process for helping understand how frequent such unusual activity occurs.}

\section{Conclusions}

Reported in this work are multiple optical photometric \& spectroscopic observations of a Be/X-ray binary system in the Magellanic Bridge undergoing a major (super-Eddington) X-ray outburst. The new data, combined with previously unpublished historical data, reveal a system that seems constantly to have a significant circumstellar disk. As a result it must always be very close to triggering an X-ray outburst. This is perhaps surprising given that the previous X-ray detection of this system was in 1993 - some 26 years ago. However, X-ray coverage of this part of the sky has been relatively sparse and future more comprehensive mapping might reveal X-ray behaviour more consistent with its optical variability, and possibly clear evidence for the binary period of this system.

\section*{Acknowledgements}

This research has made use of public MAXI data provided by RIKEN, JAXA and the MAXI team. The OGLE project has received funding from the National Science Centre,
Poland, grant MAESTRO 2014/14/A/ST9/00121 to AU.
IMM and DAHB are supported by the South African NRF. Some of the observations reported in this paper were obtained with the Southern African Large Telescope (SALT), as part of the Large Science Programme on transients 2018-2-LSP-001 (PI: Buckley). Based on observations collected at the European Organisation for Astronomical Research in the Southern Hemisphere under ESO programme 088.D-0352(A). ESB acknowledges support from the Marie Curie Actions of the European Commission (FP7-COFUND). Polish participation in SALT is funded by grant no. MNiSW DIR/WK/2016/07. This work is based on the research supported by the National Research Foundation of South Africa (Grant numbers 98969 and 93405).




\bibliographystyle{mnras}
\bibliography{references}

\label{lastpage}
\bsp	

\end{document}